# The Interactive Dance Club:
# Avoiding Chaos In A Multi Participant Environment

**Ryan Ulyate**
Synesthesia
415 South Topanga Canyon Boulevard #210
Topanga, CA 90290 USA
+1 310 455 0585
ryan@synesthesia.com

**David Bianciardi**
Synesthesia
415 Lafayette Street, 2nd floor
New York, NY 10003 USA
+1 212 353 9087
david@synesthesia.com

**ABSTRACT**
In 1998 we designed enabling technology and a venue concept that allowed several participants to influence a shared musical and visual experience. Our primary goal was to deliver musically coherent and visually satisfying results from several participants' input. The result, the Interactive Dance Club, ran for four nights at the ACM SIGGRPAH 98 convention in Orlando, Florida.

In this paper we will briefly describe the Interactive Dance Club, our "10 Commandments of Interactivity", and what we learned from it's premiere at SIGGRAPH 98.

**THE INTERACTIVE DANCE CLUB**
The Interactive Dance Club is a multi-participant interactive venue with real-time computer graphics lighting and video, synchronized to dance music (such as acid jazz, trance, ambient, drum & bass).

Instead of dancing to prerecorded music and images, members of the audience become participants. Within interactive zones located throughout the club, participants influence music, lighting, and projected imagery. There are zones for single participants, dual participants and groups. Moving from zone to zone, participants experience different blends of musical and visual elements. Like sections in an orchestra, output from the interactive zones contribute to the formation of an overall performance. Our software analyzes and filters participant's input, in order to deliver a musically coherent and visually satisfying experience.

**Design Goals**
The Interactive Dance Club is designed to:

- Allow group and individual participation in the modulation of multiple musical and computer graphics elements while maintaining a musically coherent and visually satisfying whole.
- Create a compelling social environment that amplifies the uniqueness of the individual and reveals the synergy of the group.
- Deliver to "unskilled" participants the euphoria of the artistic experience.

**Interactive Zones**
Various interfaces are located in zones throughout the club.

*Beam Breaker*
The Beam Breaker zone consists of parallel light beams above participants' heads on a dance floor with sensors that detect when a beam has been broken. Participants trigger 4-16 note musical phrases by observing when the light beams are present and "breaking" the beams with their hands.

*Infrared*
This zone consists of a state-of-the-art infrared (IR) video camera focused on an area of the dance floor. This camera registers participants' body heat. Projecting the infrared image on a large screen provides visual feedback.

*Stomp*
Stomp is a two-person zone characterized by lots of physical motion. Participants interact by dancing and stepping on 10 12"X12" floor-mounted pads (Tap Tiles, (manufactured by Infusion Systems). Discrete pad hits control short non-percussive musical phrases and projected computer graphics.

*Go-Go*
In this zone, participants step up onto a platform two feet above the dance floor and enter a cone of light. Once inside the cone, they interact by extending their arms and breaking the surrounding light beam, casting a shadow on a circular array of sensors embedded in the platform below. This controls melodic musical elements. There are 2 Go-Go zones.

*Meld Orbs*
The Meld Orbs are two four-foot-diameter spheres. Mounted on the surface of each Orb are nine near-field proximity sensors. Participants interact with each Orb to affect projected computer graphics imagery and individual notes of musical chords.

*Reach*

Reach is a one-person zone positioned on a raised platform. The participant must reach in order to touch four trigger pads positioned in a half-circle. Each pad controls an aspect of kaleidoscopic imagery and musical chords.

*Tyre-O-Mania*

Participants rotate two wheels mounted on a horizontal bar to affect projected computer graphics as well as modulate effects applied to the preprogrammed bass and drum sounds. A pair of foot pedals allows the participant to change which aspects of the computer graphics and music the wheels are controlling,

*Tweak*

Tweak is a one-person zone in which a participant interacts by swaying his or her hips between two infrared proximity sensors (D-Beams, manufactured by Interactive Light) The participant also breaks 4 light beams with their hands resulting in the triggering and modulating of percussion loops and projected computer graphics. There are 2 Tweak Zones

*Voyeur*

Participants sit on couches and manipulate remote control cameras located throughout the club. Each of the 2 Voyeur zones has 2 remote controlled cameras and 2 rear projection displays, encouraging collaboration between participants

**OUR "10 COMMANDMENTS OF INTERACTIVITY"**

To realize the Interactive Dance Club, several volunteers, in many different cities, were enlisted. Given a compressed schedule, and minimal opportunities for face to face meetings with the full group before arriving on site, much of the work had to be done in parallel. We created the following guiding principles to keep everyone on track.

**#1: Interfaces and content should encourage and reward movement**

No QWERTY keyboards, no mice! Allow unencumbered interaction. A dance club is geared towards unrestricted movement, social interaction and spontaneity. Don't use restrictive, cumbersome or isolating interfaces involving wires, gloves, goggles, etc. Content designed for the interfaces should encourage the participants to dance.

**#2: Participant's actions get an immediate and identifiable response**

No participant should ever ask "am I controlling this, or not?" Interfaces and content should respond to the participant with the same level of feedback as an automobile responds to a driver.

**#3: No instructions**

Learning to "work" the interactive zones must be intuitive and simple. There should be adequate enough feedback to for the participant to intuit if she is doing it "wrong" or "right".

**#4 People don't need to be experts to participate**

Participants are encouraged to drop their inhibitions and have fun. Nothing should be designed that intimidates people into feeling they are not good enough to participate. The system can, of course, offer deeper interaction for those that want to go further.

**#5 No thinking allowed**

The goal is to keep the participants in their "body" and not in their "head". Like a jazz musician, or a dancer, euphoria occurs when the participant gets lost in the moment, focusing on their intuitive nature. Game-like behavior causes participants to focus on their analytical side and is not appropriate in this artistic context.

**#6 Actions get aesthetically coherent responses**

Participants should navigate through and affect several "good" choices. Ensure that all participants' actions cause meaningful responses in the context of the overall performance.

**#7 Keep it simple, immediate and fun**

How long could participants do it without getting bored? Usually simpler is better. Think "Pong"

**#8 Responsiveness is more important than resolution**

In computer graphics this translates to "more speed is better than more polygons". A simple visual object that reacts quickly to participant input is better than a complex visual object that reacts too slowly.

**#9 Think modular**

Everything is a component.

**#10 Just do it... on time!**

The project will never be "finished", so hit your deadline with whatever you have ready. Then watch what it does and watch what the people do with it!

WHAT WE LEARNED

The Interactive Dance Club provided an opportunity to test out several ideas. Here are some of the things we learned.

**Interfaces**

Interfaces which allow for more freedom of movement (such as those incorporating infrared sensors) are more compatible with a dancing participant than are target driven interfaces (such as fixed, floor mounted pads). Predictable interfaces encourage predictable behavior, no matter what they actually control; if it looks like a drum, users will bang on it, even if a caress was what you'd hoped for. Interfaces with an unfamiliar appearance encouraged serendipitous interactions; if participants didn't know what to think, they didn't, and just jumped in! People will work the interfaces as hard and fast as possible; unbreakable construction and heavily filtered data were the order of the day. Based on what we learned are currently moving away from physical manipulation interfaces, and focusing our development on IR-aided machine vision systems.

**Control Systems**
Distributed, scalable systems are preferable; a little localized intelligence can go a long way towards decreasing design, programming and troubleshooting complexity.

The MIDI (Musical Instrument Digital Interface) protocol can be stretched a long way, but in the end, a TCP/IP (Transport Control Protocol / Internet Protocol) LAN (Local Area Network) with enough spare bandwidth to minimize latency is a more useful and flexible, and pretty much all protocols can be packetized and transported over IP. Top level processes need dedicated control surfaces; just like participants get user interfaces customized for their use. Experience Jockeys need dedicated controls, not mice and keyboards. By creating a balance between freedom and constraint, like a well designed game, our experiences should give enough freedom to encourage real exploration, while focusing the user. Based on what we've learned, we are currently developing a solid, scalable operating system for interactive environments.

**Music**
Contemporary dance music has a strong rhythmic component. To achieve our desire for cohesiveness we did not let any participant alter the basic drum and bass elements of the music. We chose to let participants add (or "overdub") compatible sounds on top of this basic structure. This did not diminish the participants' experience, but enhanced it by giving them a solid foundation on which to build upon. Popular music also has an identifiable beginning, middle and end. The traditional verse-chorus-bridge structure applies varying levels of musical intensity to a performance within a fixed time frame. This structure is an important element that we wanted to introduce into an interactive environment. To accomplish this, we created the role of the "Experience Jockey" (E.J.) The E.J. was able to react to the crowd and vary the overall structure in real time, providing a meta layer of direction over the entire environment. By applying this structure to the interactivity, we gave the participants a familiar framework to experience something new.

Some participants did manage, however, make a mess out of the most carefully chosen palate of compatible sounds! It's difficult to get participants to listen like musicians and react to what their neighbors are doing. There is always the tendency by some to make as much noise as fast as possible.

Not surprisingly, some of the simplest things are the most satisfying. Simply tracking participants' hip movements to modulate a band pass filter on a percussion samples was successful. This literally "hip" wah-wah effect was the source of much fun for the participants (and their friends).

**Visuals**
Real time computer graphics (CG) were done with a custom version of Side Effects Software's Houdini 3D animation software. Interfaces that we configured for trigger impulses (like drum pads and the floor mounted Tap Tiles) could not anticipate the participant gestures until they were hit, and the trigger was transmitted. This was not a problem for musical events, but it made visual events (such as a CG figure jumping when a pad is struck) look delayed and out of sync. We need to make more graphical events that correspond to the attack envelope of percussive sounds, and we need design interfaces that can give us data before a participant hits the trigger. In other words, we need to know not just when the bat strikes the baseball, but when the batter starts to swing. A lot of what we did visually was too "slow" We are developing visual content that corresponds to the percussive nature of contemporary music.

**Audience Reaction**
People shared interfaces that were designed for one (such as the Reach Zone), manipulating different parts of a one-user zone in unintended ways. People are impatient and want instant gratification; experiences with too much story or plot required more attention span than participants were willing to devote. People made up their own games with the IR zone; given some basic tools and a context, folks will self-organize fulfilling activities autonomously. People like a range of involvement from passive to active. The inclusion of couch-potato Voyeur zones, gave users remote control of pan-tilt-zoom video cams focused on the action, allowing less involved interaction.

While SIGGRAPH attracts technology savvy attendees, our venue was also patronized by attendees of a neighboring aerobic dance convention, giving us an interesting mix of mostly 18-40 year olds. Judging from the overall audience reaction, we have proceeded with technological and business development of the Interactive Club concept. We believe that the public is ready for such venues and are pursuing the goal of licensing Interactive Clubs and related technology worldwide.

**ACKNOWLEDGMENTS**
We wish to thank ACM SIGGRAPH and Walt Bransford (SIGGRAPH 98 Chair) for their support. Thanks to Judith Crow, Alan Kapler, Greg Hermanovic, Peter Wyngaard, Paul Simpson, Kristen Stratton and the other SIGGRAPH 98 volunteers as well as the 22 companies that provided equipment and support for the Interactive Dance Club

**FOR FURTHER INFORMATION**

www.synesthesia.com

info@synesthesia.com